\documentclass[%
 reprint,
 amsmath,amssymb,
 aps,
]{revtex4-2}

\usepackage{graphicx}
\usepackage{dcolumn}
\usepackage{bm}

\newcommand{\podd}{$\cal P$-odd~}
\newcommand{\todd}{$\cal T$-odd~}

\newcommand{\cm}{cm$^{-1}$}

\begin{document}

\title{Accurate \textit{ab initio} calculations of RaF electronic structure indicate the need for more laser-spectroscopical measurements} 

\author{Andrei Zaitsevskii}
\email{zaitsevskii\_av@pnpi.nrcki.ru}
\affiliation{%
 NRC ''Kurchatov Institute'' - PNPI, Orlova Roscha, 1, 188300 Gatchina, Russia  \\
 Chemistry Dept., M. Lomonosov Moscow State University, Moscow 119991, Russia 
}%

\author{Leonid V. Skripnikov}
\email{skripnikov\_lv@pnpi.nrcki.ru,leonidos239@gmail.com}
\homepage{http://www.qchem.pnpi.spb.ru}
\affiliation{
 NRC ''Kurchatov Institute'' - PNPI, Orlova Roscha, 1, 188300 Gatchina, Russia \\
 Saint Petersburg State University, 7/9 Universitetskaya nab., St. Petersburg, 199034 Russia
}%

\author{Nikolai S. Mosyagin}
\email{mosyagin\_ns@pnpi.nrcki.ru}
\affiliation{
 NRC ''Kurchatov Institute'' - PNPI, Orlova Roscha, 1, 188300 Gatchina, Russia 
}%
\author{Timur Isaev}%
 \email{timur.isaev.cacn@gmail.com, isaev\_ta@pnpi.nrcki.ru}
\affiliation{
 NRC ''Kurchatov Institute'' - PNPI, Orlova Roscha, 1, 188300 Gatchina, Russia 
}%

\author{Robert Berger}
\email{robert.berger@uni-marburg.de}
\affiliation{%
 Fachbereich Chemie, Philipps-Universit{\"a}t Marburg, Hans-Meerwein-Str 4, 35032 Marburg, Germany
}%

\author{Alexander A. Breier}
\email{a.breier@physik.uni-kassel.de}
\affiliation{%
Laboratory for Astrophysics, Institute of Physics, University of Kassel, 34132 Kassel, Germany
}%

\author{Thomas F. Giesen}
\email{t.giesen@uni-kassel.de}
\affiliation{%
Laboratory for Astrophysics, Institute of Physics, University of Kassel, 34132 Kassel, Germany
}%
\pacs{31.50.-x, 33.20.Tp, 33.70.Ca, 37.10.Mn}

\date{\today}

\begin{abstract}
Recently a breakthrough has been achieved in laser-spectroscopic studies of short-lived radioactive compounds with the first measurements of the radium monofluoride molecule (RaF) UV/vis spectra. We report results from high accuracy \emph{ab initio} calculations of the RaF electronic structure for ground and 
low-lying
excited electronic states. Two different methods agree excellently with experimental excitation energies from the electronic ground state to the $^2\Pi_{1/2}$ and $^2\Pi_{3/2}$ states, but lead consistently and unambiguously to deviations from experimental-based adiabatic transition energy estimates for the $^2\Sigma_{1/2}$ excited electronic state and show that more measurements are needed to clarify spectroscopic assignment of the $^2\Delta$ states.

\end{abstract}

\maketitle


\section{Introduction}

It is recognised that molecules with heavy nuclei are versatile tools 
to study fundamental symmetries of physical laws and the interactions
and properties of subatomic particles \cite{DeMille:08, acme:2018,
Cairncross:2017, Safronova:2018}. In such molecules
 effects resulting from both parity violation
(\podd) and time-reversal violation (\todd) can be considerably enhanced
with respect to atomic systems \cite{Flambaum:2019, Isaev:2014, Safronova:2018, kozlov:1995}. The molecules of radium monofluoride (RaF) containing different isotopes of Ra nuclei are predicted to have very high sensitivity for effects that are \podd or simultaneously \podd and \todd \cite{Isaev:2010, Isaev:2013,Borschevsky:13, Isaev:2014, Kudashov:14, Sudip:2016b, Petrov:2020} and recently an experimental breakthrough has been achieved in laser-spectroscopic studies of RaF  \cite{GarciaRuiz:2020, Udrescu:2021}. It was found that experimental values for adiabatic transition energies $T_\mathrm{e}$ and harmonic vibrational wavenumbers $\tilde{\omega}_\mathrm{e}$ for the ground and the first excited electronic states are consistent within the claimed theoretical accuracy of values reported in Refs.~\cite{Isaev:2010, Isaev:2013, Kudashov:14}. Recently extended calculations of transition energies to
low-lying electronic states of 
RaF 
have been performed in \cite{Osika:2022}, using a number of basis sets and methods available in the {\sc dirac} program package, with the authors of Ref.~\cite{Osika:2022} claiming good agreement of their theoretical values with the reported experimental data. 

The accuracy for the predicted $T_\mathrm{e}$ in RaF was estimated conservatively as 1200 \cm\ in paper \cite{Isaev:2013}, although accompanying calculations of the same quality for BaF, that could be compared to available experimental data, allowed to suggest that 
this accuracy can be at the level $\sim$500 \cm. One can expect that a large part of the uncertainty of $T_{\rm e}$ in all the mentioned calculations (including \cite{Osika:2022}) arises from the neglect of correlations involving the $5d$ shell of Ra; the uncertainty can be larger than 
that from the $4d$ shell
for BaF because of a stronger secondary relativistic destabilization of $5d$ levels. Taking into account the Ra$^+$F$^-$-like electronic structure of RaF at the equilibrium internuclear distance and the localization of low-energy excitation on the Ra$^+$, the uncertainty in excitation energies can be roughly estimated from the corresponding uncertainty for the Ra$^+$ atomic ion, which exceeds 500 cm$^{-1}$ for $7s-6d$ and 400 cm$^{-1}$ for $7s-7p$  (see Supplementary materials for details). Nevertheless, for some transitions a reasonable agreement with experimental data is still achieved due to partial error compensation
from the lack of correlations involving the $5d$ shell and 
insufficient basis set flexibility (especially in what concerns the description of angular correlations in the outer core region essential for the stabilization of $6d$-like states). 

Accuracy of {\it ab initio}
electronic structure calculations of heavy-atom compounds 
is constantly increasing
\cite{Oleynichenko:2020, Knecht:2018, Skripnikov:2020, Zaitsevskii:2020,Mosyagin:2020,Skripnikov:2021a,Skripnikov:16b}, together with
growing power of available supercomputers. For small molecules, the theoretical accuracy of calculated
 electronic transition energies approaches the level of 100 \cm~$hc$ and for certain cases even better \cite{Skripnikov:2021a,Skripnikov:2021b}. In Ref.~\cite{Skripnikov:2021a} a method to take into account contributions of quantum electrodynamics (QED) effects to transition energies in molecules has been proposed and implemented. This formulation of the model QED Hamiltonian is closely related to the formulation of model QED operator in Ref.~\cite{Shabaev:13}, that is now widely used for atomic calculations~\cite{Shabaev:13,Schwerdtfeger:2017,Kaygorodov:2021}.

Thus it is timely to perform precise calculations of the electronic structure of RaF, providing transition energy estimates with an uncertainty well below the values of vibrational quanta. To assess the reliability of the values of obtained molecular parameters 
we use two different 
schemes to compute spectroscopic parameters for the ground and excited electronic states. Special care is taken to study systematically the uncertainties introduced by different approximations. 

\section{Calculation methods}

\subsection{Scheme 1: Fock-space relativistic coupled cluster  calculations}

The present 
Fock-space relativistic coupled cluster (FS RCC)
excited-state studies were performed using two basic electronic structure models. One series of calculations employed  accurate shape-consistent relativistic pseudopotentials (RPP) derived from the valence solutions of atomic four-component Dirac--Fock--Breit equations with Fermi nuclear charge distributions~\cite{Mosyagin:10a,Mosyagin:16,ourGRECP}. The Ra pseudopotenial replaced the inner core shells $1\!-\!4s$, $2\!-\!4p$, $3\!-\!4d$, $4f$; relativistic and finite-nuclear-size effects for the fluorine atom were described by the `empty-core' pseudopotential leaving all electrons for explicit treatment~\cite{Mosyagin:21}. Alternatively, a Dirac--Coulomb--Gaunt Hamiltonian was used to solve the closed-shell SCF problem and then converted to the two-component all-electron Hamiltonian by means of the X2C technique within the molecular mean field approximation (X2C MMF, \cite{Sikkema:09}).   

The employed FS RCC scheme of correlation treatment closely resembles that used in our previous study on RaCl~\cite{Isaev:21}. The Fermi vacuum was defined by the ground-state determinant of the positive molecular ion, while the target neutral states were considered as belonging to the one-particle $(0h1p)$ sector.  The FS RCC active space normally comprised 9 Kramers pairs of lowest-energy virtual spinors of RaF$^+$ arising from the $7s$, $7p$ and $6d$ spinors of Ra$^+$. Electronic transition energies as functions of the internuclear separation $R$ were evaluated within the singles-and-doubles approximation for the cluster operator (FS RCCSD). In most cases, 7 electrons of F and 19 electrons of Ra$^{+}$ (including the $5d$ shell) are correlated.  The $[10s\, 9p\, 9d\, 7f\, 4g\, 3h\, 2i]$ Ra basis set compatible with the RPP model is taken from Ref.~\cite{Isaev:21}; its essential feature consists in using ANO-type high-angular-momentum $(g,\, h,\, i)$ functions optimized for Ra and Ra$^+$ within the scalar relativistic approximation. Its counterpart for all-electron calculations was obtained by combining the same $[4g\, 3h\, 2i]$ ANOs and diffuse $f$ functions with the primitive $(28s\, 25p\, 18d\, 12f)$ Gaussian sets from Ref.~\cite{Roos:05}. The fluorine basis used with both models of relativistic Hamiltonians was the aug-cc-pvQZ one~\cite{Woon:93} adapted to the relativistic treatment~\cite{Jong:01}.

\begin{figure}
\begin{center}
\includegraphics[width=0.95\columnwidth]{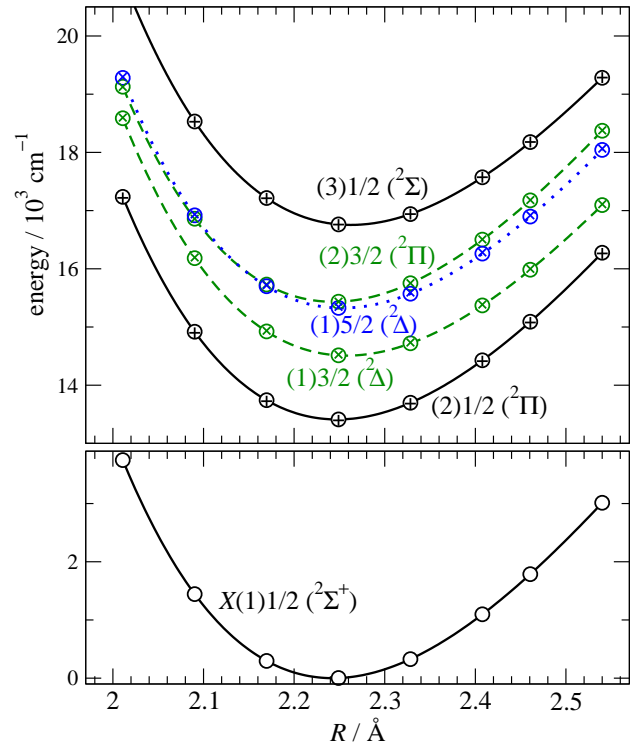}
\caption{\label{PotCurves} Potential energy functions for low-lying electronic states of RaF. Excited state functions are derived from
FS RCC excitation energies 
computed within the pseudopotential model (curves and empty circles) and all-electron X2C MMF approximation (crosses).}
\end{center}\end{figure}

Excited-state potential energy curves were constructed by adding the FS RCCSD electronic transition energies as functions of the internuclear separation $R$ to the accurate ground state potential which was computed by means of the single-reference coupled cluster method with perturbative account for the contribution from triple excitations (RCCSD(T) scheme) and counterpoise corrections of the basis set superposition errors (cf. Ref.~\cite{Pazyuk:15}). 
Calculated potential energy curves are given in Fig.~\ref{PotCurves}.

The contributions from the correlations involving the core $5s5p$ subshells of Ra were estimated in a single-point ($R$=2.249 \AA) FS RCCSD calculations performed with an appropriately modified basis: the $[4g\, 3h\, 2i]$ ANO set of Ra was replaced by $[5g\, 4h\, 3i]$ ANOs optimized to describe the correlations of all explicitly treated electrons of Ra. One can also expect certain contributions arising from higher-rank terms in the cluster operator expansion. Unfortunately, full nonperturbative treatment even of connected triple excitations (FS RCCSDT) remains unfeasible whereas the Fock space analogues of efficient single-reference schemes with perturbative triples (like the famous CCSD(T) one) are not reliable~\cite{Oleynichenko:2020}. To estimate the effect of connected triples on the computed excitation energies,  we computed FS RCCSD and FS RCCSDT effective Hamiltonians ($H^{\rm eff}_{\rm SD,\, r}$ and $H^{\rm eff}_{\rm SDT,\, r}$ respectively) correlating only 17 electrons of RaF and rejecting the cluster operator components involving one-electron levels above a certain threshold (up to 2.2~$E_\mathrm{h}$). Approximate FS CCSDT energies are then obtained as eigenvalues of the operator    
\[
H^{\rm eff}_{\rm SDT}\approx H^{\rm eff}_{\rm SD,\, full}+H^{\rm eff}_{\rm SDT,\, r}-H^{\rm eff}_{\rm SD,\, r}.
\] 
where $H^{\rm eff}_{\rm SD,\, full}$ is the FS RCCSD effective Hamiltonian calculated with no restriction imposed on single and double excitations in the cluster operator. The $T_{\rm e}$ values incorporating the resulting corrections for triple excitations and $5s5p$ correlations are marked as ``final'' in Table \ref{alldata}.

The composition of relativistic states in terms of their scalar relativistic counterparts was determined by the projection technique described in Ref.~\cite{Zaitsevskii:17}. The requisite scalar relativistic states were obtained within the same computational scheme by switching off the spin-orbit parts of the pseudopotentials. Since the projection analysis was restricted to the model-space parts of the wavefunctions, it was considered reasonable to extend the model space for this task, augmenting the number of active spinor pairs to 34. To suppress the effect of intruder states normally encountered for large model spaces, the technique of simulated imaginary shifts of energy denominators~\cite{Oleynichenko:20cpl} was employed.

The construction of one-electron spinors and molecular integral evaluation, as well as the single-reference RCCSD(T) ground state calculations, were performed with the DIRAC 19 code \cite{DIRAC:19,DIRAC:20} whereas the EXP-T program \cite{EXPT:20,Oleynichenko:2020} was used for FS RCC calculations. Vibrational energy levels were evaluated with the help of the program VIBROT \cite{Sundholm}.

\subsection{Scheme 2: single-reference calculations}
Another series of calculations employed the single reference approach as a base. For this, we have followed the scheme which was developed in Ref.~\cite{Skripnikov:2021a} (see also~\cite{Skripnikov:2021b}) and applied to calculation of excitation energies for low-lying electronic states of Ra$^+$ and transition energy of the first excited state of RaF. This scheme included the following steps: The main correlation calculation has been performed within the CCSD(T) method using the Dirac-Coulomb Hamiltonian. All electrons were included in the correlation treatment and the virtual energy cutoff has been set to 10000~$E_\mathrm{h}$. Such cutoff ensures that correlation contributions of the inner-core electrons are described correctly~\cite{Skripnikov:17a,Skripnikov:15a} which is important for the case of all-electron calculation. The basis set for Ra optimized in Ref.~\cite{Skripnikov:2021a} was used. It corresponds to the modified uncontracted Dyall's AEQZ~\cite{Dyall:12} basis set augmented by diffuse functions of $s$-, $p$-, $d$-, and $f$-types. Functions of $g$-, $h$-, and $i$- types were partly replaced by uncontracted natural-like functions constructed using the procedure and code developed in Refs.~\cite{Skripnikov:2020e,Skripnikov:13a,Skripnikov:16a}. In total, the basis set for Ra included $[42s\, 38p\, 27d\, 17f\, 11g\, 3h\, 2i]$ functions. The uncontracted AETZ~\cite{Dyall:12} basis set has been used for F. Contribution of the Gaunt interelectron interaction to transition energies of RaF has been treated at the FS-CCSD level within the X2C MMF approach~\cite{Sikkema:09}. To consider the basis set extension contribution, the basis set on F has been increased up to the uncontracted AAEQZ~\cite{Dyall:12} one and basis set on Ra up to $[42s\, 38p\, 27d\, 27f\, 13g\, 9h\, 6i]$~\cite{Skripnikov:2021a}. This basis set extension contribution has been calculated within the FS-CCSD method with excluded $1s\ldots 3d$ electrons of Ra. To take into account more functions with L$\le$6 we have performed  calculations within the 37-electron EOM-EA-CCSD~\cite{Nooijen:1995} approach using the scalar-relativistic variant of the RPP operator~\cite{Titov:99,Mosyagin:10a,Mosyagin:16,ourGRECP}. Such approximation has been tested in Ref.~\cite{Skripnikov:2021a} (see also~\cite{Skripnikov:16b}). In particular, such an approach allowed us to take into account the contribution of $[15g\, 15h\, 15i]$ functions which was practically impossible within the Dirac-Coulomb calculations. Following Ref.~\cite{Skripnikov:2021a} we have also added extrapolated correction on harmonics with L$>$6. The contribution of iterative triple and perturbative quadruple cluster amplitudes has been obtained within the CCSDT(Q) method~\cite{Kallay:6} using the two-component RPP Hamiltonian~\cite{Titov:99,Mosyagin:10a,Mosyagin:16,ourGRECP}. The basis set consisting of natural compact contracted $(20s\, 20p\, 15d\, 10f)/[6s\, 6p\, 7d\, 4f]$ functions~\cite{Skripnikov:13a,Skripnikov:16b,Skripnikov:2020e} has been used for Ra, while the aug-cc-pvDZ-DK basis set~\cite{Kendall:1992,Jong:01} has been employed for F. 
In the correlation calculation on the CCSDT(Q) level, 27 outer electrons of RaF have been included and the virtual energy cutoff has been set to 5~$E_\mathrm{h}$. Finally, we have calculated the contribution of the vacuum-polarization and self-energy quantum-electrodynamics effects to the electronic energies of molecular terms. For vacuum polarization operator we have used the model Uehling potential approximate formula from Ref.~\cite{Ginges:2016}. The self-energy contribution has been calculated within the model QED Hamiltonian using the expression suggested and developed in Ref.~\cite{Skripnikov:2021a}. This formulation is close to the expression developed and applied for atomic calculations in Ref.~\cite{Shabaev:13}. 

Four-component calculations have been performed within the {\sc dirac} code \cite{DIRAC:19,DIRAC:20}. High order correlation effects were calculated using the~{\sc mrcc}~\cite{MRCC2020,Kallay:1} code. Scalar relativistic correlation calculations to ensure the basis set completeness and to generate compact basis sets were performed using the {\sc cfour}~\cite{CFOUR} code. The code developed in Ref.~\cite{Skripnikov:2021a} has been employed to calculate the QED contribution to molecular and atomic transition energies.

\section{Results and discussion}
In presented calculations of the electronic structure of RaF we 
analyzed 
the main sources of possible 
theoretical uncertainties.
Within the scheme 1, systematic errors  are primarily due to the basis set incompleteness with certain contributions  from  incomplete account for triples and the neglect of 
higher cluster amplitudes
in the cluster operator as well as the neglect of QED effects. Since the electronic structure of RaF in all states under study roughly correspond to ionic configurations, Ra$^+$F$^-$, it seems reasonable to derive the corrections from the comparison of computed excitation energies at the Ra$^+$ + F$^-$ dissociation limit with the experimental data on Ra$^+$. The \emph{ab initio}  energies, corresponding to $7s-6d$ and $7s-7p$ excitations of the free Ra$^+$, are systematically overestimated (by 125 cm$^{-1}$ to 235 cm$^{-1}$). Unfortunately, as follows from rather large transition moments to the ground state arising mainly from the $7s$ state of Ra$^+$  (Table \ref{alldata}), low-lying molecular states which can be formally associated with the $6d$ states of Ra$^+$ receive a significant $7p$ contribution, so that the common practice of shifting each potential curve to fit exactly the corresponding experimental dissociation limit seems not well founded.  However, one can hope to improve $T_{\rm e}$ estimates  by shifting uniformly all excited-state energies to minimize the overall error for all $6d$ and $7p$ limits. The corresponding $T_{\rm e}$ values are referred to as ``shifted'' in Table \ref{alldata}.

As in Ref.~\cite{Skripnikov:2021a} the main uncertainty of scheme 2 is the remaining basis set incompleteness, neglect of the retardation part of the Breit interaction and interference between different contributions such as high order correlation effects and contribution of high harmonics in the basis set. The theoretical uncertainty of the prediction within scheme 2 is expected to be about 5 meV (40 \cm) as in Ref.~\cite{Skripnikov:2021a}. The resulting energies of scheme 2 and  the ``shifted'' $T_{\rm e}$ energies obtained within scheme 1 are in perfect agreement. Note, that we did not apply any empirical energy shifts to molecular transition energies within scheme 2. 

The energies of the excited electronic states of the BaF molecule corresponding to states considered here for RaF were studied in Ref.~\cite{Skripnikov:2021b}.
Application of the calculation approach similar to scheme 2~\cite{Skripnikov:2021a} to BaF resulted in the agreement of all considered theoretical transition energies with the experimental ones within $\approx$20 \cm~\cite{Skripnikov:2021b}.

It can be seen from Tables \ref{alldata} and \ref{composition} that the first excited electronic 
state can be reliably identified as $^2\Pi_{1/2}$ and the theoretical and experimental values for 
$T_\mathrm{e}$ are in excellent agreement. The next state ($^2\Delta_{3/2}$) has to be
 in the region of 14000 \cm, a wavenumber region that was not investigated in the experiment \cite{GarciaRuiz:2020} though. The state at $\sim$15100 \cm\, assigned tentatively in \cite{GarciaRuiz:2020} as $^2\Delta_{3/2}$, is rather of $^2\Delta_{5/2}$ type and transitions to this state from the ground electronic state are possible due to rovibronic coupling effects
(estimates in ~\cite{Petrov:19private} suggested $\sim$0.05\% transition intensity of $  X \rightarrow\:^2\Delta_{5/2}$ transition in units of $ X  \rightarrow\:^2\Delta_{3/2}$ transition intensity).
For the next $^2\Pi_{3/2}$ state 
we again see excellent agreement with the experiment, but for the next $^2\Sigma_{1/2}$ state theory and experiment disagree
on the level of 450 \cm. On purely energetical grounds, a possible explanation might be an incorrect vibrational quanta assignment of the electronic transitions into the 
excited
$^2\Sigma_{1/2}$ state starting instead from a vibrational hot level of the electronic ground state into the vibrational ground state of the excited state. Re-evaluation of the four observed transitions into the 
excited
$^2\Sigma_{1/2}$ state \cite{GarciaRuiz:2020}, assuming them, now, as hot band transitions ($ X(\nu^{\prime\prime}) \rightarrow\: ^2\Sigma_{1/2}(\nu^{\prime}): 1\rightarrow 0, 2\rightarrow 1, 3\rightarrow 2, \text{ and } 4\rightarrow 3$), leads to an experimental $T_{\rm e}$ value of 16620.8(2)\,cm$^\text{-1}$, agreeing excellently with the  theoretical value. The observed vibronic profile for this transition would, however, be at variance with the Franck--Condon profile expected by virtue of the small change in equilibrium distance upon electronic excitation. New experimental RaF-studies in the spectral region around 16620\,cm$^\text{-1}$ might, thus, provide valuable information about the fundamental transition from $ X \rightarrow\: ^2\Sigma_{1/2}$, confirming the assignment of vibrational quanta.

The main difference of the high-accuracy quantum chemical calculations reported in the present work as compared to our earlier theoretical predictions reported in Ref.~\cite{Isaev:2013}, however, pertains to the systematic energetical lowering of the $^2\Delta$ manifold of states by about 650\,\cm, suggesting the low-lying $^2\Delta_{3/2}$ state to be as of yet experimentally unidentified. This level remains energetically well above the lowest excited $^2\Pi_{1/2}$ state as predicted earlier when studying the prospects for laser-coolability of RaF \cite{Isaev:2010,Isaev:2013}.

\begin{table*}[h]\caption{\label{alldata} FS RCC molecular constants for the low-lying electronic states of RaF. The meaning of abbreviations are as follow: RPP - calculations within the pseudopotential model; AE - all-electron X2C MMF calculations;   
 $\Delta T({\rm T})$ - estimated contribution of triple cluster amplitudes and $\Delta T(5s5p)$ - contributions from correlations of core-like $5s5p$ subshells to relative term energies. The equilibrium distances $R_e$ are in \AA, adiabatic transition energies $T_\mathrm{e}$ and harmonic vibrational quanta $\omega_{\rm e}$ are in \cm, FS RCC model-space estimates of squared transition dipole moments 
 to the ground electronic state $d^2$ are in $e^2\,a_0^2$.
}
\begin{tabular*}{0.9\textwidth}{@{\extracolsep{\fill}}lccccc}
\hline
\hline 
State           &  (2)1/2  & (1)3/2  & (1)5/2 & (2)3/2  & (3)1/2 \\
                &  ($^2\Pi_{1/2}$)  & ($^2\Delta_{3/2}$)  & ($^2\Delta_{5/2}$) & ($^2\Pi_{3/2}$)  & ($^2\Sigma_{1/2}$) \\
\hline 
\multicolumn{6}{c}{Molecular parameters, scheme 1}\\
$R_e$ (RPP)         & 2.248 & 2.258 &  2.253 & 2.243 & 2.263 \\ 
\hspace{2.4ex} (AE)        & 2.247 &2.258 & 2.253 & 2.243 & 2.262 \\
 \\ 
$\omega_{\rm e}$ (RPP)        & 436.8 & 430.3 & 433.8 & 436.3 & 431.9 \\
\hspace{2.4ex} (AE)     & 436.9 & 430.8 & 433.9 & 436.4 & 432.1 \\
\hspace{2.4ex} (Exptl)\cite{GarciaRuiz:2020}  & 435.5 &       &       &       &       \\
\\
\multicolumn{6}{c}{Transition energies, scheme 1:} \\
$T_{\rm e}$ (RPP)         & 13412 & 14509 &  15327 & 15434 & 16754 \\
\hspace{2.4ex} (AE, 27$e$)&  13396 &14522 &  15345 &15435 &  16751 \\
\\
$\Delta T({\rm T})$       & +49   &  $-86$ & $-74$ &  +58  &  +16     \\                            
$\Delta T(5s5p)$       & +23   &  +62 & +62 &  +38  &  +30     \\
\\
$T_{\rm e}$(RPP, final)     &  13484  &  14485 & 15315  &  15530 & 16800 \\ 
\hspace{2.4ex} (RPP, final, shifted by $-164~\mathrm{cm}^{-1}$)     & \bf  \bf 13320 & \bf 14321 & \bf 15151  & \bf 15366 &\bf 16636 \\  
$d^2$ (RPP) &7.8&0.33& 0 & 7.0 & 8.4 \\  
\\
\multicolumn{6}{c}{Transition energies, scheme 2:} \\
CCSD(T), DC, 97e              & 13381    & 14603   & 15402 & 15463  & 16746   \\
High harmonics, CBS (L)       &        2 &   -98   &   -98 &     1  &   -15   \\
CCSDT(Q)-CCSD(T), 2c-RPP, 27e &      -28 &   -15   &    -5 &   -40  &   -26   \\
Gaunt                         & 5        &   -65   &   -78 &   -18  &   -11   \\
QED                           &      -56 &   -73   &   -70 &   -57  &   -50   \\
$T_e$, final, scheme2         &    {\bf 13303}~\cite{Skripnikov:2021a} & \bf{14352}   & \bf{15151} & \bf{15348}  & \bf{16644}   \\
\\ 
$T_{\rm e}$(Exptl) \cite{GarciaRuiz:2020}            &  13288 &  15148(?)&           & 15355 &  16181\\
&  &  &           & &  16620.8(2)\footnote{When re-assigned, for details see text.}\\
\\ 
                   
\hline 
\hline
\multicolumn{6}{c}{Data from \cite{Isaev:2013}} \\
$T_e$, AE(DC), 17$e$, FS-CCSD/$10^4$ &1.33 & 1.50 & 1.58 & 1.54& 1.67 \\
\multicolumn{6}{c}{Data from \cite{Osika:2022}} \\
$T_e$, RPP, 19$e$, FS-CCSD &13298 & 14978 & 15740 & 15332& 16614 \\
\hline
\hline 
\end{tabular*}
\\
Ground state:  $R_e$=2.244 \AA, $\omega_e$=440.6 cm$^{-1}$ (exptl 441.8(1) cm$^{-1}$)
\end{table*}

\begin{table}[h]\caption{\label{composition}Composition of full relativistic states of RaF ($R$(Ra--F)=4.25\,$a_0$) in terms of scalar relativistic states.}
\begin{center}
\begin{tabular}{rl}
\hline 
\hline \\ 
$X(1) 1/2$  &  100\% $(1)^2\Sigma^+$  \\
$ (2) 1/2$  &  86\% $(1)^2\Pi$, 13\% $(2)^2\Sigma^+$ \\
$ (1) 3/2$  &    96\% $(1)^2\Delta$, 4\% $(1)^2\Pi$ \\
$ (1) 5/2$  &    100\% $(1)^2\Delta$ \\
$ (2) 3/2$  &    96\% $(1)^2\Pi$, 3\% $(1)^2\Delta$ \\
$(3)1/2  $  &   87\% $(2)^2\Sigma^+$, 13\% $(1)^2\Pi$ \\
\\ \hline 
\hline 
\end{tabular}
\end{center}
\end{table} 

\section{Conclusion}
We have calculated molecular parameters and transition frequencies between the ground and five low-lying excited electronic states of the RaF molecule on a new level of accuracy. We used two different high-accuracy calculation schemes and achieved good agreement between these two theoretical studies, but not with experiment for the adiabatic transition energy from the electronic ground to the excited $^2\Sigma_{1/2}$ state, whereas excellent agreement between theory and experiment is found for transitions to the states of approximate $^2\Pi_{1/2}$ and $^2\Pi_{3/2}$ character. Our results also indicate that more spectroscopic measurements are needed to clarify the spectroscopic assignment of the $^2\Delta_{3/2}$ state.

\section{Acknowledgement}
The authors are grateful to A.N. Petrov for early estimates of non-adiabatic effects for $^2\Delta_{5/2}$ state. 

 This work has been carried out using computing resources of the federal collective usage centre Complex for Simulation and Data Processing for Mega-science Facilities at NRC ``Kurchatov Institute'', http://ckp.nrcki.ru/, and computers of Quantum Chemistry Lab at NRC ``Kurchatov Institute" - PNPI.
 
 L.V.S.\ acknowledges support by the Russian Science Foundation under Grant No. 19-72-10019 for the electronic structure calculations within the scheme 2 and RFBR according to research project No. 20-32-70177 for the calculations of QED and Gaunt contributions. 
 The work on the RPP generation for the light elements was supported by the personal scientific fellowship for N.S.M.\ from the governor of Leningrad district.
 T.I. and A.Z. are grateful for financial support of FS RCC studies to RSF-DFG grant N 21-42-04411.

\bibliographystyle{apsrev}

\end{document}